\documentclass[aps,prl,twocolumn,showpacs,superscriptaddress]{revtex4-1}
\usepackage{graphicx}
\usepackage{epstopdf}
\usepackage{amssymb}
\usepackage{ulem}
\usepackage{xcolor}

\begin{document}

\title{Superconducting gap and vortex lattice of the heavy fermion compound $\mathrm{CeCu_2Si_2}$}
\author{Mostafa Enayat}
\altaffiliation{These authors contributed equally.}
\affiliation{Max-Planck-Institut f\"ur Festk\"orperforschung, Heisenbergstrasse 1, D-70569 Stuttgart, Germany}
\author{Zhixiang Sun}
\altaffiliation{These authors contributed equally.}
\affiliation{Max-Planck-Institut f\"ur Festk\"orperforschung, Heisenbergstrasse 1, D-70569 Stuttgart, Germany}
\author{Ana Maldonado}
\altaffiliation{These authors contributed equally.}
\affiliation{Max-Planck-Institut f\"ur Festk\"orperforschung, Heisenbergstrasse 1, D-70569 Stuttgart, Germany}
\affiliation{SUPA, School of Physics and Astronomy, University of St. Andrews, North Haugh, St. Andrews, Fife, KY16 9SS, United Kingdom}
\author{Hermann Suderow}
\affiliation{Laboratorio de Bajas Temperaturas, Departamento de F\'isica de la Materia Condensada, Instituto de Ciencia de Materiales Nicol\'as Cabrera, Condensed Matter Physics Center (IFIMAC), Facultad de Ciencias Universidad Aut\'onoma de Madrid, 28049 Madrid, Spain}
\affiliation{Unidad Asociada de Bajas Temperaturas y Altos Campos Magn\'eticos, UAM, CSIC, Cantoblanco, E-28049 Madrid, Spain}
\author{Silvia Seiro}
\affiliation{Max-Planck-Institut f\"ur die Chemische Physik fester Stoffe, N\"othnitzer Strasse 40, 01187 Dresden, Germany}
\author{Christoph Geibel}
\affiliation{Max-Planck-Institut f\"ur die Chemische Physik fester Stoffe, N\"othnitzer Strasse 40, 01187 Dresden, Germany}
\author{Steffen Wirth}
\affiliation{Max-Planck-Institut f\"ur die Chemische Physik fester Stoffe, N\"othnitzer Strasse 40, 01187 Dresden, Germany}
\author{Frank Steglich}
\affiliation{Max-Planck-Institut f\"ur die Chemische Physik fester Stoffe, N\"othnitzer Strasse 40, 01187 Dresden, Germany}
\author{Peter Wahl}
\email{wahl@st-andrews.ac.uk}
\affiliation{SUPA, School of Physics and Astronomy, University of St. Andrews, North Haugh, St. Andrews, Fife, KY16 9SS, United Kingdom}
\affiliation{Max-Planck-Institut f\"ur Festk\"orperforschung, Heisenbergstrasse 1, D-70569 Stuttgart, Germany}

\date{\today}
\begin{abstract}
The order parameter and pairing mechanism for superconductivity in heavy fermion compounds are still poorly understood. Scanning tunneling microscopy and spectroscopy at ultra-low temperatures can yield important information about the superconducting order parameter and the gap structure. Here, we study the first heavy fermion superconductor, CeCu$_2$Si$_2$. Our data show the superconducting gap which is not fully formed and exhibits features that point to a multi-gap order parameter. Spatial mapping of the zero bias conductance in magnetic field reveals the vortex lattice, which allows us to unequivocally link the observed conductance gap to superconductivity in CeCu$_2$Si$_2$. The vortex lattice is found to be predominantly triangular with distortions at fields close to $\sim 0.7 H_{c2}$.
\end{abstract}


\maketitle

Superconductivity in heavy fermion materials was first observed in CeCu$_2$Si$_2$ \cite{Steglich1979} and was unexpected. The formation of heavy fermion bands is usually ascribed to the interaction between delocalized conduction electrons and localized magnetic moments, whereas localized magnetic moments in conventional superconductors rapidly destroy superconductivity. Since this discovery, superconductivity has been found in a range of other heavy fermion materials, often in close proximity to a quantum phase transition between a magnetically ordered phase and a phase dominated by Kondo screening. This proximity of superconductivity to a magnetic quantum critical point found in many Ce-based compounds \cite{mathur1998} indicates that magnetic fluctuations and the influence of the quantum critical point might promote superconductivity in these materials. The physics near the quantum critical point sensitively depends on the balance between screening of the local magnetic moments and interactions between them\cite{Doniach}. In CeCu$_2$Si$_2$, slight changes in the exact composition (specifically the Cu-to-Si ratio) result in superconducting crystals ($S$ type), magnetically ordered samples ($A$ type) or samples which exhibit competing phases ($A/S$ type) \cite{Feyerherm,seiro2010}. Superconductivity occurs close to a spin density wave-type quantum critical point and is expected to be unconventional in nature. Neutron scattering data indicate that superconductivity is mediated by spin fluctuations rather than by phonons \cite{stockert2011}. Despite intense research, the precise form of the superconducting order parameter remains elusive. Measurements of the relaxation rate of the nuclear quadrupolar resonance \cite{Ishida, Fujiwara} and specific heat under pressure \cite{Lengyel} have yielded evidence for an unconventional order parameter. Angle-resolved resistivity measurements of the upper critical field $H_{\mathrm c2}$ exhibit a fourfold symmetry of $H_{\mathrm c2}$ consistent with a $d_{xy}$ symmetry of the order parameter \cite{vieyra2011}. In contrast, recent thermodynamic measurements offer evidence for nodeless multiband superconductivity in $\mathrm{Ce}\mathrm{Cu}_2\mathrm{Si}_2$, challenging the view that the pairing symmetry is of nodal $d$-wave type \cite{Kittaka2014}. Also from theory, different symmetries of the order parameters and different coupling mechanisms have been proposed, e.g. a $d$-wave symmetry for superconductivity mediated by magnetic fluctuations \cite{Eremin2008} and, very recently, $s_{\pm}$-wave symmetry emerging from magnetic fluctuations of higher order\cite{Ikeda2015}.

Scanning tunneling microscopy and spectroscopy has been successfully employed to study heavy fermion materials\cite{schmidt2010, Aynajian2010} as well as Kondo lattice compounds\cite{Ernst2011, Wahl2011}. Yet, due to the low temperatures required for these experiments, STM and STS have only been applied to few heavy fermion superconductors successfully \cite{sakata2000,suderow2004,yazdani2012,Maldonado2012,Allan2013,Zhou2013}, despite these having been thoroughly studied using macroscopic techniques (see, e.g., ref.~\cite{Pfleiderer} for a review). In particular, quasiparticle interference imaging provides evidence for a $d_{x^2-y^2}$ pairing symmetry in CeCoIn$_5$ \cite{Allan2013,Zhou2013}. Imaging the vortex lattice and transitions in the vortex lattice arrangement can yield additional insight into the structure of the superconducting gap \cite{Zhou2013}. Here we use a dilution refrigerator-based STM to study the superconducting properties of CeCu$_2$Si$_2$ at temperatures down to 20 mK \cite{udai2013}. Experiments have been performed on an $S$ type single crystal with a superconducting transition temperature $T_\mathrm c = 0.59 \mathrm K$ and an upper critical field along $c$ of $\mu_{0}H_\mathrm{c2} \sim 2.3\mathrm T$.% and $H_{c2}(0) = 1.96 \mathrm T$.

\begin{figure}
\centering
\includegraphics[width=0.45\textwidth]{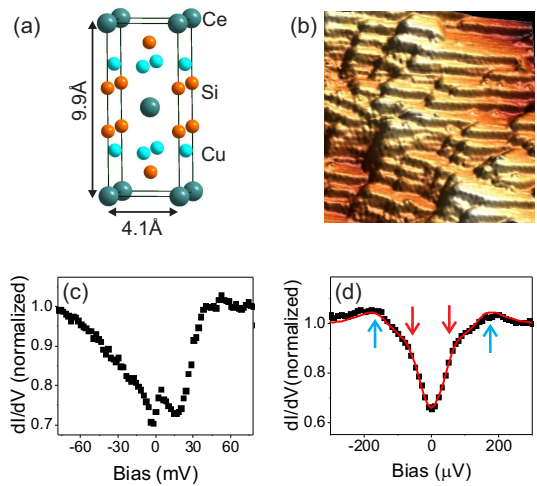}
\caption{(a) Body centered tetragonal crystal structure of CeCu$_2$Si$_2$. (b) Topographic image of the surface after cleavage showing terraces which exhibit atomic periodicities and step heights of $\sim 2\mathrm{\AA}$ (30$\times$30 nm$^{2}$, $T=4.5\mathrm K$), (c) high energy normalized tunneling spectrum taken at $T=200\mathrm{mK}$ showing a W-shaped feature close to the Fermi energy ($V_\mathrm{RMS}=3\mathrm{mV}$). (d) Normalized tunneling spectrum recorded at $20\mathrm{mK}$ ($V_\mathrm{RMS}=40\mathrm{\mu V}$), exhibiting a gap-like feature with two shoulders symmetric with respect to zero bias. Blue arrows indicate the main gap, red arrows the low energy shoulder. Red line shows the best fit obtained for two gaps, one of $d$-wave symmetry at low energies and one of $s$-wave symmetry at high energies (for details see text and Ref.~\onlinecite{epaps}).}
%topo: run mk11\cecu2si2\ topo 132
\label{topo}
\end{figure}

CeCu$_2$Si$_2$ crystallizes in the $\mathrm{Th}\mathrm{Cr}_2\mathrm{Si}_2$ $I$4/$mmm$ body-centered tetragonal structure with lattice parameters $a = 4.1031(2)\mathrm\AA$ and $c = 9.9266(5) \mathrm\AA$ \cite{seiro2010}, see Fig.~\ref{topo}(a). Many materials with the same crystal structure cleave easily perpendicular to the crystallographic $c$-axis (e.g. the 122 iron pnictides, YbRh$_2$Si$_2$, CeRh$_2$Si$_2$, YbCo$_2$Si$_2$ and URu$_2$Si$_2$), due to differences in the bond strengths favoring breakage of the crystal within specific planes. CeCu$_2$Si$_2$ crystals do not cleave in a way which yields extended, atomically flat terraces. The bonds within the unit cell appear to be of rather similar strength which prevents the development of a preferred cleaving plane\cite{cleavenote}. Out of more than thirty single crystals which we cleaved, only two resulted in images which exhibit atomic periodicity and atomically flat terraces as reported below.

To promote cleaving in the $ab$ plane, a groove was cut around the perimeter of the sample in a plane perpendicular to the $c$ axis. In order to avoid surface contamination the sample was cleaved in-situ in cryogenic vacuum at a cleaving stage fixed to the $4\mathrm K$ plate of the cryostat. After cleaving, the sample was immediately transferred into the STM head. Differential conductance spectra were measured with a lock-in amplifier, with a bias modulation of between 10 and 40 $\mu\mathrm V_\mathrm{RMS}$ (unless stated otherwise). The base temperature of the instrument is below $20\mathrm{mK}$, the electronic temperature of the instrument has been determined previously to be $137\mathrm{mK}$ \cite{udai2013}.

After successful cleavage of the sample, flat regions in an area of several hundred nanometers were found. In the absence of a natural cleavage plane, different cleaves result in different surface terminations, and also across the sample, the termination is not the same everywhere. Fig.~\ref{topo}(b) shows an example of atomically flat terraces with steps of a height of $2\mathrm\pm 0.2\mathrm\AA$ in between. For the specific area shown in fig.~\ref{topo}(b), we can analyze the atomic periodicity and tilt angle and can attribute it to a (105) surface. Spectra taken in a range of $\pm80\mathrm{mV}$ show a W-shaped feature close to the Fermi level(see Fig.~\ref{topo}(c)), reproducible across different surface terminations. It is suggestive to interpret the peak-like feature near zero bias as a signature of the heavy 4f-bands, its position is consistent with the energy of the Kondo resonance determined from angular resolved photoemission\cite{Ehm2007}.

At the lowest temperature achieved, $20\mathrm{mK}$, the differential tunneling conductance measured on this surface reveals a gap-like feature on an energy scale of $100\mathrm{\mu eV}$ which we attribute to the superconducting gap, see Fig.~\ref{topo}(d). The spectra do not show a strong dependence on the location on the surface. The tunneling spectrum of the superconducting gap exhibits a rather complex structure with weak coherence peaks at $\pm 170\mu\mathrm{V}$, additional shoulders at $\pm75\mu\mathrm V$ and a substantial zero bias conductance of about $65\%$ of the normal state differential conductance. We have attempted to describe the data by different models for the superconducting gap (for details of the fits see \cite{epaps}). We find an excellent agreement for two gaps, one which describes the low energy features by a superconducting gap with a $d$-wave symmetry and a second one describing the high energy features by a gap with $s$-wave symmetry. The extracted fitting parameters give $\Delta_{1}=(56\pm3)\mu\mathrm{eV}$ for the $d$-wave gap and $\Delta_{2}=(147\pm4)\mu\mathrm{eV}$ for the $s$-wave gap. These findings point towards a rather complex multi-gap superconducting order parameter. The ratio of the gap magnitudes of the large and the small gap obtained from the best fit to the tunneling data $\Delta_{2}/\Delta_1\approx 2.6$ compares well with the ratio of 2.5 deduced from specific heat \cite{Kittaka2014}. The shape of the tunneling spectra is similar to the ones observed in CeCoIn$_{5}$\cite{Zhou2013,Allan2013}.

\begin{figure}
\centering
\includegraphics[width=0.5\textwidth]{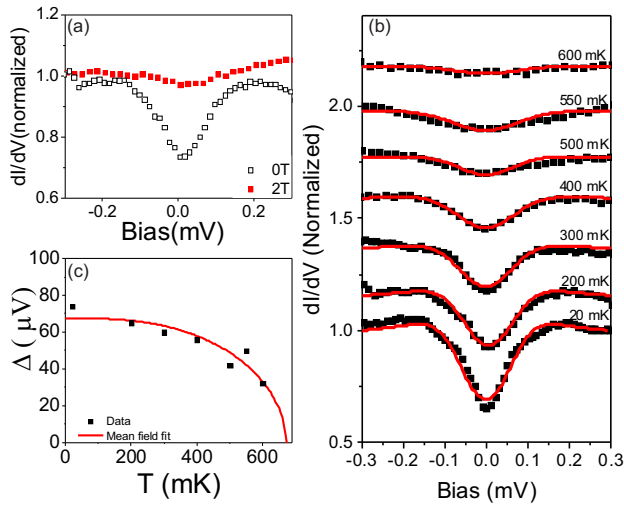}
\caption{(a) Differential conductance spectra (normalized at $-0.4\mathrm{mV}$ recorded at $200\mathrm{mK}$ in zero field (open black symbols) and in a field of $2\mathrm T$ (solid red symbols).
(b) Scanning tunneling spectroscopy as a function of temperature ($V_\mathrm{RMS}=24\mathrm{\mu V}$, except at 20mK $V_\mathrm{RMS}=40\mathrm{\mu V}$). Red lines represent fits with the Dynes equation for a single $s$-wave gap \cite{Dynes1978}; spectra have been shifted vertically for clarity. (c) Gap size as function of temperature extracted from fits in (b), the red line represents the mean field behaviour expected for an $s$-wave order parameter, $\Delta(T)=\Delta_{0}\tanh(\frac{\pi}{2}\sqrt{\frac{T_\mathrm c}{T}-1})$. The fit yields $\Delta_0=67\pm3\mathrm{\mu eV}$ and $T_\mathrm c=670\pm30\mathrm{mK}$. }
\label{spectra}
\end{figure}

Fig.~\ref{spectra}(a) shows for comparison tunneling spectra taken at $T=200\mathrm{mK}$ ($<T_\mathrm c$) at zero magnetic field and at a field of $\mu_0 H=2\mathrm T$ parallel to $c$, close to the upper critical field $H_{\mathrm c2}$ \cite{seiro2010}. As can be seen from the spectrum measured at $2\mathrm T$, the gap has almost closed completely. The temperature dependence of the spectra as shown in Fig.~\ref{spectra}(b) reveals that the gap vanishes as the temperature rises from $20\mathrm{mK}$ to $600\mathrm{mK}$.

To assess the temperature dependence of the gap size in a more quantitative way, we have used the Dynes equation \cite{Dynes1978} to fit the spectra in Fig.~\ref{spectra}(b) with a single isotropic $s$-wave gap. While a multi-gap order parameter would be more appropriate, the fit becomes unstable as the features of the spectra become broadened at higher temperatures. The broadening parameter $\Gamma$ has been fixed to its value at base temperature, $\Gamma = 71\mu\mathrm{eV}$ and is likely dominated by gap anisotropy and the multigap structure. In addition we have accounted for broadening due to the finite temperature and resolution. While the fits yield overall good agreement with the data, there are also clear deviations, especially for the spectra obtained at the lowest temperatures. We attribute both the fit quality and the large broadening parameter to the fact that the superconducting order parameter in $\mathrm{CeCu}_2\mathrm{Si}_2$ is more complex than just a single $s$-wave gap.
We expect the characteristic gap size extracted in this way to be qualitatively representative of the behaviour of the true gap. The temperature dependence of the gap size $\Delta(T)$ obtained in this way is shown in Fig.~\ref{spectra}(c). It reveals similarity to a mean-field BCS behaviour (red solid line) \cite{ruckenstein1987mean}. Extrapolation of the gap size to zero Kelvin (see fig.~\ref{spectra}(c)) yields a value of $67\pm3\mathrm{\mu eV}$, slightly smaller than what is found from an isotropic $s$-wave fit at base temperature ($\Delta = 71 \mu\mathrm{eV}$ at $20\mathrm{mK}$).
%******************* move to Discussion **************************
The difference of the temperature dependence of the gap size for $s$-wave and $d$-wave order parameters is rather minor, so that we would not expect to be able to distinguish them from the temperature dependence of the gap size\cite{won1994}.

\begin{figure}
\centering
\includegraphics [width=0.5\textwidth]{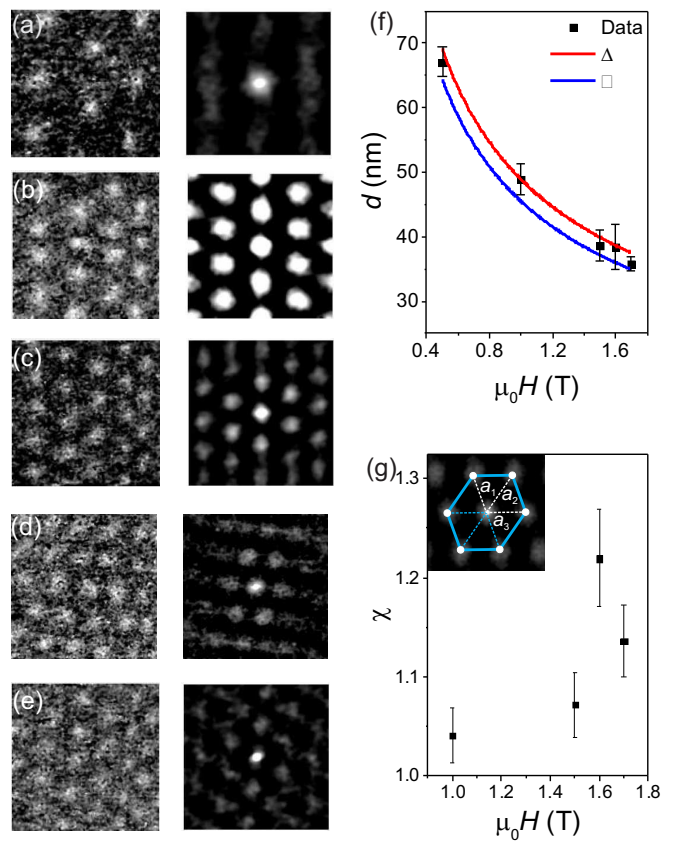}
\caption{ (a-e) Maps of the zero bias tunneling conductance (ZBTC) at 0.2 K (150$\times$150 nm$^{2}$, $V_\mathrm{RMS}=40\mathrm{\mu V}$) taken after cooling the sample in zero field as a function of magnetic field (0.5, 1, 1.5, 1.6 and 1.7 T), together with the autocorrelation. The data have been filtered for better contrast. The triangular vortex lattice is clearly seen. (f) Distance between the vortex cores as a function of magnetic field. The points are obtained from averaging the distances in the three high symmetry directions from the autocorrelation. Solid lines show the expected dependencies for a triangular (red line) and square lattice (blue line).
%$d_\triangle=(\frac{2}{\sqrt{3}}\frac{\Phi_0}{H})^{\frac{1}{2}}$, $d_\Box=(\frac{\Phi_0}{H})^{\frac{1}{2}}$ respectively.
(g) Anisotropy of the vortex lattice extracted from the autocorrelation of differential conductance maps as a function of magnetic field. As a measure for the anisotropy, we use $\chi$ (see eq.~\ref{chidef}, inset shows the anisotropy exemplarily for the autocorrelation of the map taken at $1.6\mathrm T$). Maps taken with a tunneling set point of $U=5\mathrm{mV}$ (except (d) with $U=3\mathrm{mV}$), $I=0.6\mathrm{nA}$.}
\label{vortex}
%a: 812 (0.5T), b: 805 (1T), c: 759 (1.5T), d: 846 (1.6T), e: 799 (1.7T)
%812: 5mV, 0.6nA, 77nm/s, 180x180pix, 3layers, mod: 50µV
%805: 5mV, 0.6nA, 80nm/s, 190x190pix, 15layers, mod: 50µV
%759: 5mV, 0.6nA, 63nm/s, 190x190pix, 15layers, mod: 50µV
%846: 3mV, 0.6nA, 23nm/s, 256x256pix, 20layers; dito 883: -1.5mV, 0.8nA, 10nm/s, 150x150pix, 20layers, mod: both 50µV
%also 849, 814
%799: 5mV, 0.6nA, 80nm/s, 190x190pix, 15layers, mod: 50µV
%not shown:
%748 (2T)
\end{figure}

Spatial maps of the zero bias tunneling conductance (ZBTC) at finite magnetic fields can reveal the vortex lattice\cite{Hess1989}. Observation of the vortex lattice can provide further information on the order parameter. For isotropic order parameter and electronic structure, the vortices usually arrange themselves in a triangular Abrikosov lattice\cite{blatter2004}. Other symmetries of the vortex lattice can occur if there is an anisotropic interaction between the vortex cores, e.g. due to the symmetry of the superconducting order parameter or anisotropies of the electronic structure \cite{tokuyasu1990vortex,xu1996structures,ichioka1996vortex}.
Maps of the differential conductance at zero bias were taken at magnetic fields of $\mu_0 H=0.5, 1, 1.5, 1.6, 1.7 \mathrm T$, see Fig.~\ref{vortex}(a to e). As expected, the number of vortices is proportional to the applied magnetic field. Despite the rather rough surface morphology, our data show an ordered vortex lattice which is close to triangular symmetry.

We have analyzed the distance between the vortex cores $a_i$ from the autocorrelation of the ZBTC maps along the high symmetry directions of the vortex lattice. The average distance between the neighboring vortices as a function of magnetic field reveals a behaviour consistent with a triangular lattice (fig.~\ref{vortex}(f)). As can be seen from figs.~\ref{vortex}(a)-(e), our data indicate a distortion of the vortex lattice away from regular triangular symmetry, which becomes largest at magnetic fields of $1.6\mathrm T$. At this field, the vortex lattice also appears with a different orientation than at smaller or larger fields. We define
\begin{equation}
\chi=\mathrm{max}(a_1,a_2,a_3)/\mathrm{min}(a_1,a_2,a_3)
\label{chidef}
 \end{equation}
as a measure for the asymmetry of the vortex lattice, with $\chi=1$ corresponding to a regular triangular vortex lattice (see inset in fig.~\ref{vortex}(g)). The magnetic field dependence of $\chi$ is plotted in fig.~\ref{vortex}(g), confirming a significant distortion of the vortex lattice specifically at a field of $1.6\mathrm T$\cite{comment05}. This distortion is robust across multiple data sets taken with different parameters, excluding drift as a possible reason. Also the surface roughness is unlikely the cause for the anisotropy, because this is similar for the images in fig.~\ref{vortex}(a-e).
%The surface is substantially tilted with respect to the crystallographic $c$-axis, as can be seen from the high density of step edges. Overall, it is interesting to note that the vortex lattice appears not to be influenced by the surface morphology, forming a well ordered lattice despite the high defect density at the surface.
%*******Discussion***********

\begin{figure}
\centering
\includegraphics[width=0.5\textwidth]{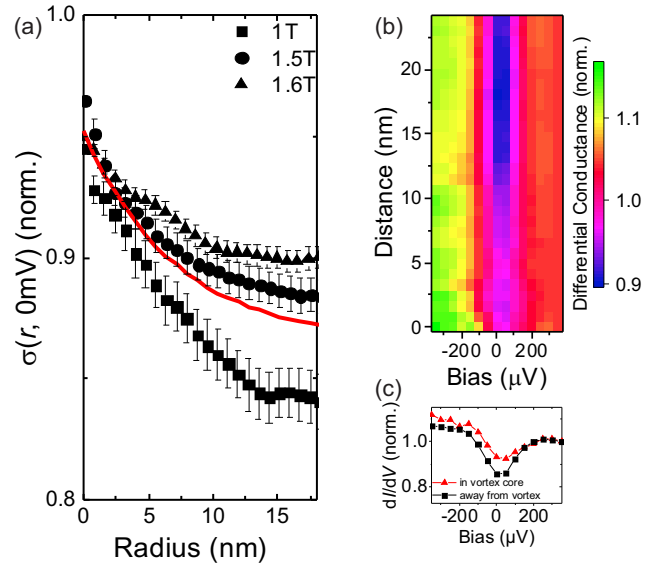}
\caption{
(a) Evolution of the azimuthal average of the normalized ZBTC $\sigma(r,0)$ at 0.2 K and 1T, 1.5T and 1.6T as a function of the radial distance $r$ from the vortex center (error bars represent the standard deviation of the data, average over multiple vortex cores). From fits of Eq.~\ref{eq:vortex10} to the individual curves we obtain $\xi= 7.1\pm 0.3\mathrm{nm}$ on average (solid line).
(b) Azimuthally averaged tunneling spectra as a function of distance from a vortex core measured at $1\mathrm T$. The zero bias tunneling conductance is substantially larger inside the vortex core than away from the vortex core (data has been median filtered for clarity, spectra normalized at a bias voltage of $350\mu\mathrm V$).
(c) tunneling spectra from (b) in the vortex core (red triangles) and away from a vortex core (black squares). Even in the vortex core, the zero bias tunneling conductance stays below the normal state value (i.e. at larger bias voltages).
}
\label{analysis}
\end{figure}

Further insight into the properties of superconductivity can be obtained by analyzing individual vortex cores. In fig.~\ref{analysis}(a), we show the decrease of the zero bias conductance, $\sigma(r,0)$, as a function of distance $r$ from the center of the vortex core, as superconductivity recovers. The characteristic length scale $\xi$ of this recovery, which is a measure of the superconducting coherence length, is determined by fitting an exponential decay to the radial profile of the ZBTC \cite{Pan2000},
\begin{equation}
\sigma(r,0)=\sigma_0\cdot e^{-r/\xi} + c,
\label{eq:vortex10}
\end{equation}
where $\sigma_0$ is the additional conductance in the vortex core and $c$ accounts for a constant background.
%see also PRB43, 7609
We obtain $\xi=7.1\pm 0.3\mathrm{nm}$ averaging the decay lengths obtained from the data shown in fig~\ref{analysis}(a). This value of $\xi$ is close to the one obtained previously for the coherence length ($10\mathrm{nm}$)\cite{Rauchschwalbe1982}.
%compare 5.6nm to Hc2

For superconductors with an isotropic gap and which are in the clean limit, strong vortex bound states are expected near the Fermi level \cite{Caroli1964}, which give rise to a strong peak in differential conductance spectra recorded in the center of a vortex core\cite{Hess1989}. Spectra of the differential conductance acquired inside a vortex core and away from it show no evidence for a vortex bound state (see Fig.~\ref{analysis}(b, c)). This behaviour is consistent with a not fully formed gap on some bands or a considerable amount of scattering in the superconductor (or at the surface). The mean free path in $\mathrm{CeCu}_2\mathrm{Si}_2$ is on the order of $10\mathrm{nm}$\cite{Rauchschwalbe1982}, therefore the material is not in the clean limit.\\
The zero bias conductance extracted from tunneling spectra at zero field agrees reasonably well with specific heat data of Kittaka {\it et al.} \cite{Kittaka2014}: at zero field and $200\mathrm{mK}$, our tunneling spectra yield a ZBTC of $0.74\pm 0.05$ of the normal state conductance, whereas the normalized specific heat at $200\mathrm{mK}$ ($C_s/C_n$) is about $0.65$ of the one in the normal state. This shows that the superconducting gap we detect in tunneling spectra at the surface is consistent with what one would expect from bulk superconductivity as measured by specific heat\cite{tsutsumi2014}.

It is interesting to compare our observations to the case of $\mathrm{CeCoIn}_5$, where experiments suggest a $d$-wave symmetry of the order parameter\cite{Allan2013,Zhou2013}. The spectra we have observed at base temperature (see fig.~\ref{topo}(d)) show a striking similarity with tunneling spectra of the superconducting gap in the heavy fermion superconductor CeCoIn$_5$ \cite{Allan2013,Zhou2013}.  Small angle neutron scattering of the vortex lattice in $\mathrm{CeCoIn}_5$ has previously shown substantial variations from triangular to square and rhombic lattice symmetries\cite{Bianchi2008}.

In conclusion, our tunneling spectroscopic measurements on the first discovered heavy fermion superconductor, CeCu$_2$Si$_2$, show clear evidence for superconductivity with at least two distinct gaps, with the best fits obtained for the smaller with $d$-wave character and the larger with $s$-wave character. Although the observed gap does not open fully, the temperature and magnetic field dependent measurements establish its link to superconductivity. Characterization of the vortex lattice shows a triangular vortex lattice at low fields with significant distortion at fields around $0.7H_{c2}$ indicating an anisotropic interaction between vortex cores. The features observed in the tunneling spectra at low energy support a multi-gap scenario. The shape of the gap detected in tunneling spectroscopy as well as the distortion of the vortex lattice indicate similarities with superconductivity in CeCoIn$_5$.

We gratefully acknowledge K. Machida and Y. Fasano for stimulating discussions. ZS acknowledges support from NWO (grant no. 680.50.1119).


\begin{thebibliography}{99}

%\bibitem{Fisk1986} Z. Fisk, H.R. Ott, T.M. Rice, J.L. Smith, {\it Nature} {\bf 320}, 124-129, (1986).
\bibitem{Steglich1979} F. Steglich, J. Aarts, C. D. Bredl, W. Lieke, D. Meschede, W. Franz, H. Sch\"afer, {\it Phys. Rev. Lett.} {\bf 43}, 1892, (1979).

%\bibitem{Bardeen} J. Bardeen, L.N. Cooper and J.R. Schrieffer, Phys. Rev. {\bf 108}, 1175 (1957).
\bibitem{mathur1998} N.D. Mathur, F.M. Grosche, S.R. Julian, I.R. Walker, D.M. Freye, R.K.W. Haselwimmer and G.G. Lonzarich, {\it Nature} {\bf 394}, 39 (1998).

\bibitem{Doniach} S. Doniach, Physica B {\bf 91}, 231 (1977).

\bibitem{Feyerherm} R. Feyerherm, {\it et al.}, Phys. Rev. B {\bf 56}, 699 (1997).

\bibitem{seiro2010} S. Seiro, M. Deppe, H. Jeevan, U. Burkhardt and C. Geibel, {\it Phys. Status Solidi B} {\bf 247}, 614--616 (2010).

\bibitem{stockert2011} O. Stockert, J. Aarndt, E. Faulhaber, C. Geibel, H.S. Jeevan, S. Kirchner, M. Loewenhaupt, K. Schmalzl, W. Schmidt, Q. Si, F. Steglich {\it  Nat. Phys. } {\bf 7}, 119, (2011).

\bibitem{Ishida} K. Ishida, {\it et al.}, Phys. Rev. Lett. {\bf 82}, 5353 (1999).

\bibitem{Fujiwara} K. Fujiwara, {\it et al.}, J. Phys. Soc. Jpn. {\bf 77}, 123711 (2008).

\bibitem{Lengyel} E. Lengyel, {\it et al.}, Phys. Rev. B {\bf 80}, 140513 (2009).

\bibitem{vieyra2011} H. A. Vieyra, N. Oeschler, S. Seiro, H. S. Jeevan, C. Geibel, D. Parker, F. Steglich, {\it Phys. Rev. Lett.} {\bf 106}, 207001, (2011).

\bibitem{Kittaka2014} S. Kittaka, Y. Aoki, Y. Shimura, T. Sakakibara, S. Seiro, C. Geibel, F. Steglich, H. Ikeda, and K. Machida {\it Phys. Rev. Lett.} {\bf 112}, 067002 (2014).

\bibitem{Eremin2008} I. Eremin, G. Zwicknagl, P. Thalmeier, and P. Fulde, {\it Phys. Rev. Lett.} {\bf 101}, 187001 (2008).

\bibitem{Ikeda2015} H. Ikeda, M.-T. Suzuki, and R. Arita {\it Phys. Rev. Lett.} {\bf 114}, 147003 (2015).

\bibitem{schmidt2010} A.R. Schmidt, M. H. Hamidian, P. Wahl, F. Meier, A. V. Balatsky, J. D. Garrett, T. J.Williams, G. M. Luke and J. C. Davis, {\it Nature} {\bf 465}, 570 (2010).

\bibitem{Aynajian2010} P. Aynajian,  E.H. da Silva Neto, C. V. Parker, Y. Huang, A. Pasupathy, J. Mydosh, and A. Yazdani, {\it Proc. Natl. Acad. Sci.} {\bf 107}, 10383 (2010).

\bibitem{Ernst2011} S. Ernst, S. Kirchner, C. Krellner, C. Geibel, G. Zwicknagl, F. Steglich, and S. Wirth, {\it Nature} {\bf 474}, 362--366 (2011).

\bibitem{Wahl2011} P. Wahl, L. Diekh\"oner, M.A. Schneider, F. Treubel, C.T. Lin, and K. Kern, {\it Phys. Rev. B.} {\bf 84}, 245131 (2011).

\bibitem{yazdani2012} P. Aynajian, E. H. da Silva Neto, A. Gyenis, R. E. Baumbach, J. D. Thompson, Z. Fisk, E. D. Bauer, and A. Yazdani, {\it Nature} {\bf 486}, 201 (2012)

\bibitem{Maldonado2012} A. Maldonado, I. Guillamon, J. G. Rodrigo, H. Suderow, S. Vieira, D. Aoki, and J. Flouquet, {\it Phys. Rev. B.} {\bf 85}, 214512 (2012).

\bibitem{sakata2000} H. Sakata, N. Nishida, M. Hedo, K. Sakurai, Y. Inada, Y. Onuki, E. Yamamoto, Y. Haga, {\it J. Phys. Soc. J.} {\bf69}, 1970--1973, (2000).

\bibitem{suderow2004} H. Suderow, S. Vieira, J.D. Strand, S. Bud'ko, P.C. Canfield, {\it Phys. Rev. B} {\bf 69}, 060504, (2004).

\bibitem{Zhou2013} B. B. Zhou, S. Misra, E. H. da Silva Neto, P. Aynajian, R. E. Baumbach, J. D. Thompson, E. D. Bauer and A. Yazdani, {\it  Nat. Phys.} {\bf 9}, 474--479, (2013).

\bibitem{Allan2013} M. P. Allan, F. Massee, D. K. Morr, J. Van Dyke, A. W. Rost, A. P. Mackenzie, C. Petrovic and J. C. Davis, {\it  Nat. Phys.} {\bf 9}, 468--473, (2013).

\bibitem{Pfleiderer} C. Pfleiderer, {\it Rev. Mod. Phys.} {\bf 81}, 1551--1624 (2009).

\bibitem{udai2013} U. Singh, M. Enayat, S. White, and P. Wahl, {\it Rev. Sci. Instrum.} {\bf 84}, 013708 (2013).

\bibitem{cleavenote} This is likely related to the comparatively small c/a ratio, which indicates a so-called collapsed structure with strong Si-Si bonds between adjacent Cu-Si layers.

\bibitem{Ehm2007} D. Ehm, S. H\"ufner, F. Reinert, J. Kroha, P. W\"olfle, O. Stockert, C. Geibel, and H. v. L\"ohneysen, {\it Phys. Rev. B} {\bf 76}, 045117 (2007).

\bibitem{epaps} see supplementary information in the EPAPS.

\bibitem{Dynes1978} R.C. Dynes, V. Narayanamurti, and J. P. Garno, {\it Phys. Rev. Lett.} {\bf 41}, 1509 (1978).

\bibitem{ruckenstein1987mean} Ruckenstein, A. E. and Hirschfeld, P. J. and Appel, J., {\it Phys. Rev. B.} {\bf 36}, 857 (1987).

\bibitem{won1994} H. Won and K. Maki, $d$-wave superconductor as a model of high-$T_\mathrm{c}$ superconductors, {\it Phys. Rev. B} {\bf 49}, 1397 (1994).

\bibitem{Hess1989} H.F. Hess, R.B. Robinson, R.C. Dynes, J.M. Valles, and J.V. Waszczak, {\it Phys. Rev. Lett.} {\bf 62}, 214 (1989).

\bibitem{blatter2004} G. Blatter, M. V. Feigelman, V. B. Geshkenbein, A. I. Larkin, V. M. Vinokur {\it Rev. Mod. Phys.} {\bf 66}, 1125 (1994).

\bibitem{tokuyasu1990vortex} Tokuyasu, T.A. and Hess, D.W. and Sauls, J.A., {\it Phys. Rev. B.}, {\bf 41}, 8891 (1990),

\bibitem{xu1996structures} Ji-Hai Xu and Yong Ren and Chin-Sen Ting, {\it Phys. Rev. B.}, {\bf 53}, R2991 (1996),

\bibitem{ichioka1996vortex} Ichioka, M. and Hayashi, N. and Enomoto, N. and Machida, K., {\it Phys. Rev. B.}, {\bf 53}, 15316 (1996).

\bibitem{comment05} There is also a deviation at $0.5\mathrm T$, where however we observe only a few vortex cores within the field view and therefore it is difficult to extract lattice distances.

\bibitem{Pan2000} S.H. Pan, E.W. Hudson, A.K. Gupta, K.-W. Ng, H. Eisaki, S. Uchida, and J.C. Davis, {\it Phys. Rev. Lett.} {\bf 85}, 1536--1539 (2000).

\bibitem{Rauchschwalbe1982} U. Rauchschwalbe, W. Lieke, C. D. Bredl, F. Steglich, J. Aarts, K. M. Martini, A. C. Mota,  {\it Phys. Rev. Lett.} {\bf 49}, 1448--1451, (1982)

%\bibitem{Eskildsen2002} M.R. Eskildsen, M. Kugler, S. Tanaka, J. Jun, S.M. Kazakov, J. Karpinski, and O. Fischer  {\it Phys. Rev. Lett.} {\bf 89}, 187003, (2002).

\bibitem{Caroli1964} C. Caroli, P.G. de Gennes, and J. Matricon  {\it Phys. Lett.} {\bf 9}, 307--309, (1964).

\bibitem{tsutsumi2014} Y. Tsutsumi, K. Machida and M. Ichioka, {\it Phys. Rev. B} {\bf 92}, 020502 (2015).

\bibitem{Bianchi2008} A.D. Bianchi, {\it et al.}, {\it Science} {\bf 319}, 177--180 (2008).

\end{thebibliography}
\end{document}